# Experimental analysis of variability in $WS_2$-based devices for hardware security

M. Vatalaro[1], H. Neill[2], F. Gity[2], P. Magnone[3], V. Maccaronio[1], C. Márquez[4], J. C. Galdon[4], F. Gamiz[4], F. Crupi[1], P. Hurley[2], R. De Rose[1,*]

[1]*DIMES, University of Calabria, Rende 87036, Italy*
[2]*Nanoelectronic Materials and Devices Group, Tyndall National Institute, University College Cork, Cork T12 R5CP, Ireland*
[3]*Department of Management and Engineering, University of Padova, Vicenza 36100, Italy*
[4]*Nanoelectronics Research Group (CITIC-UGR), Department of Electronics, University of Granada, Granada 18071, Spain*
[*]*Corresponding author. E-mail address: r.derose@dimes.unical.it*

**Abstract**

This work investigates the variability of tungsten disulfide ($WS_2$)-based devices by experimental characterization in view of possible application in the field of hardware security. To this aim, a preliminary analysis was performed by measurements across voltages and temperatures on a set of seven $Si/SiO_2/WS_2$ back-gated devices, also considering the effect of different stabilization conditions on their conductivity. Obtained results show appreciable variability in the conductivity, while also revealing similar dependence on bias and temperature across tested devices. Overall, our analysis demonstrates that $WS_2$-based devices can be potentially exploited to ensure adequate randomness and robustness against environmental variations and then used as building blocks for hardware security primitives.

## 1. Introduction

Two-dimensional (2D) materials are widely recognized as promising technology thanks to appealing electrical, optical, and mechanical characteristics, which are attracting interest in a wide variety of fields such as transistor electronics, flexible electronics, nanophotonics, etc. [1-5]. Taking advantage of their properties, several research groups have recently fabricated 2D materials-based electronic devices with remarkable performance [5]. However, the yield and reliability of such devices are largely affected by different types of intrinsic (vacancies, impurities, atomic misalignments, thickness fluctuation in the 2D sheet) and extrinsic (e.g., variable adhesion and interaction with the adjacent materials) defects and issues coming from the manufacturing process, also leading to large device-to-device variability [5]. Although the latter typically represents an undesired effect, it can be exploited for implementing hardware security primitives leveraging stochastic process variations to generate a unique, random, and secure ID, like a fingerprint [6]. Indeed, the intrinsic variability along with the mechanical flexibility of 2D materials-based devices make them particularly suitable for next-generation flexible electronics, while also targeting hardware security applications (e.g., smart labels for anticounterfeiting) [3, 7].

Within the context presented above, this work focuses on the experimental characterization of variability in a set of seven back-gated devices based on tungsten disulfide ($WS_2$), i.e., a 2D semiconducting material belonging to the transition metal dichalcogenides (TMDCs) family [8], aiming at preliminarily evaluating their suitability for hardware security primitives. More specifically, the analysis was performed with reference to a Physically Unclonable Function (PUF) bitcell implementation consisting of a voltage divider between two nominally identical series-connected devices [9, 10], as shown in Fig. 1, also illustrating a typical output voltage ($V_{OUT}$) distribution arising from $M_1$-$M_2$ mismatch. Indeed, this circuit leverages the devices mismatch in terms of their conductivity to generate a random voltage at the output node. At the same time, the devices must exhibit similar bias and temperature dependences to ensure adequate robustness of the PUF response against voltage and temperature variations, as well as a similar aging effect for response repeatability over time. To this aim, the variability of the conductivity of $WS_2$-based devices was evaluated across voltages and temperatures, while also considering the effect of different stabilization conditions.

The rest of the paper is organized as follows. Section 2 details the manufacturing process and the properties of the analyzed devices. Section 3 reports and discusses measurement results at room temperature, whereas Section 4 shows the effect of temperature on device characteristics. Finally, Section 5 summarizes the main conclusions of this work.

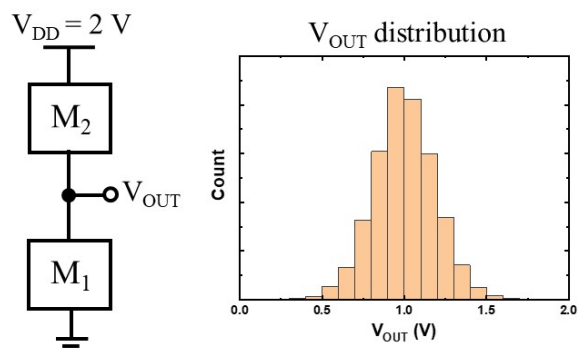

**Fig. 1.** Two-device voltage divider for PUF implementation and a typical output voltage ($V_{OUT}$) distribution arising from $M_1$-$M_2$ mismatch.

## 2. Manufacturing process and device description

Fig. 2 illustrates the sketch of the fabricated WS$_2$-based devices, consisting of a Si/SiO$_2$/WS$_2$ back-gated structure. WS$_2$ films were directly grown on an n-type Si substrate covered by a 90-nm thick thermally grown SiO$_2$ layer. First, a tungsten trioxide (WO$_3$) thin film was deposited by thermal evaporation on the Si substrate at wafer level, with a deposition rate of 2.5 Å/s. The boat was filled up with WO$_3$ powder (Aldrich 99.9) at a pressure of 8×10$^{-5}$ mbar. WS$_2$ film synthesis was accomplished via chemical vapor deposition (CVD) through sulfurization of WO$_3$ at a maximum temperature of 750 °C with a growth time of 20 minutes. One alumina crucible containing 0.4 g of sulfur powder (Aldrich 99.9) and pieces of silicon substrate (about 2.5cm×1.5cm) were placed in a quartz process tube (2-inch diameter). The quartz tube was then introduced into a two-temperature-zone furnace (Planartech) and the pressure was set to 10 Torr throughout the process. The tube was initially purged using argon gas (600 sccm) for 5 min at room temperature. The temperature was then increased to 150 ºC over 10 min in the WO$_3$ stream. The argon flow was then reduced to 400 sccm and the downstream and upstream furnaces were heated to 750 ºC and 180 ºC, respectively. Subsequently, both furnaces were turned off and were naturally cooled down to 500 ºC followed by a fast cooling to room temperature in argon ambient. Finally, back-gated devices were fabricated by patterning Ni/Au metal stack as source/drain electrodes using the metal lift-off process. Isolated devices were then patterned by fluorine-based ICP dry etching process using resist as mask, which was stripped by dipping the sample in acetone.

Fig. 3 shows the micrograph of the fabricated sample along with the Atomic Force Microscopy (AFM) analysis of one typical device with channel length $C_L = 50$ µm and channel width $C_W \approx 1.7$ µm. AFM data also show an average channel height of about 20 nm with the rms roughness of 2.68 nm (i.e., ~13%).

## 3. Measurement results at room temperature

The electrical characterization of seven devices under test (DUTs) with nominally identical dimensions ($C_L = 50$ µm and $C_W = 1.7$ µm) was performed through a Cascade SUMMIT 11861B probe station equipped with a Temptronic chuck temperature controller and a Keithley 4200-SCS parameter analyzer. The electrical characteristics were initially evaluated at room temperature ($T = 25$ °C). For a fair comparison, the transfer characteristics (drain current $I_D$ vs gate-source voltage $V_{GS}$ ranging from +35 V to -35 V and back at drain-source voltage $V_{DS} = 2$ V) of the different devices were measured after an initial stabilization phase, which consisted of applying $V_{GS} = 35$ V and $V_{DS} = 2$ V for about 470 s. To appreciate the effect of such stabilization, Fig. 4(a) reports the time evolution of $I_D$ during such a phase, whereas Fig. 4(b) shows the comparison of $I_D$-$V_{GS}$ curves obtained without and with stabilization for one typical device. From Fig. 4, we can observe that the stabilization at $V_{GS} = 35$ V leads to an increased device conductivity, while also strengthening its hysteresis behavior. The latter was quantified by calculating the $V_{GS}$ shift (i.e., $\Delta V_{GS}$) corresponding to the $I_D$ value at $V_{GS} = 0$ V during +35 V to -35 V sweep. Both effects can be mainly ascribed to charge trapping/detrapping effects due to defects near or at the oxide/channel interface [11, 12]. Fig. 4(b) also reveals a PMOS behavior, which can be presumably attributed to defects and vacancies in the WS$_2$ film inducing Fermi level pinning at the interfaces with the metal contacts and the back-gate oxide [13]. Both hysteresis and PMOS behaviors were consistently observed in all analyzed devices, as shown in Fig. 5(a) reporting the $I_D$-$V_{GS}$ curve obtained after the initial stabilization for each device. Moreover, Fig. 5(b) reports the corresponding histogram of the $\Delta V_{GS}$ shift as extracted from the $I_D$-$V_{GS}$ curves, which shows a very low variability in terms of hysteresis (i.e., $\sigma/\mu = 2.4\%$). It is worth pointing out that in general the observed hysteretic behavior (typically depending on the level of charge trapping at the oxide/channel interface) could be detrimental for the circuit performance, especially in the case the devices belonging to the same circuit are subjected to different operating conditions. However, the very low device-to-device variability in terms of hysteresis suggests that the effects of charge trapping/detrapping mechanisms are quite consistent across analyzed devices although they exhibit rather different conductivity, thus limiting the impact of such a behavior at the circuit level.

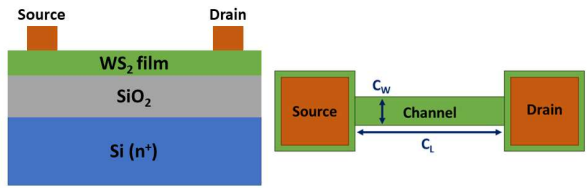

**Fig. 2.** Schematic cross section (left) and top view (right) of the WS$_2$-based device, showing channel width ($C_W$) and channel length ($C_L$).

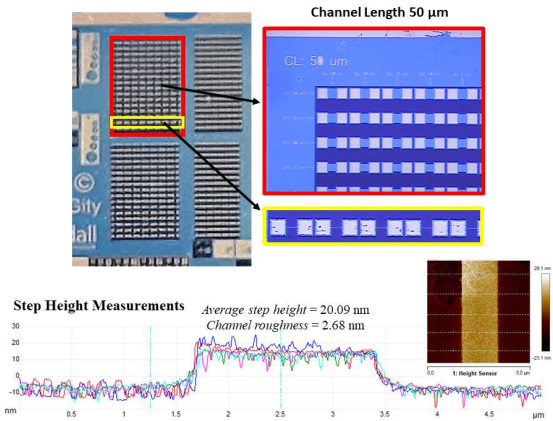

**Fig. 3.** Sample photo (top) and AFM analysis of a typical device (bottom).

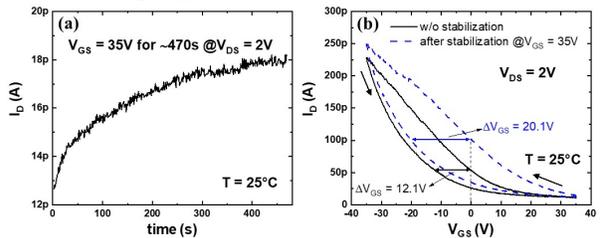

**Fig. 4.** (a) Time evolution of $I_D$ during stabilization with $V_{GS} = 35$ V and $V_{DS} = 2$ V for about 470 s and (b) $I_D$-$V_{GS}$ curves at $V_{DS} = 2$V obtained without and with stabilization for one typical device at 25 °C.

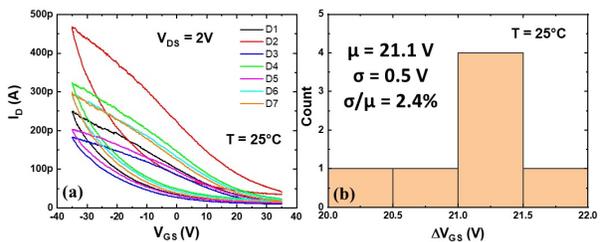

**Fig. 5.** (a) $I_D$-$V_{GS}$ curves at $V_{DS} = 2$ V after stabilization with applying $V_{GS} = 35$ V and (b) histogram of the hysteresis-induced $\Delta V_{GS}$ shift across seven devices at 25 °C.

The device-to-device variability was evaluated in more detail by measuring the output characteristics ($I_D$ vs $V_{DS}$ ranging from +2 V to -2 V and back at zero-$V_{GS}$) of DUTs, which are reported in Figs. 6(a)-(b) as obtained without any initial stabilization. Measurement results expectedly show no hysteresis in $I_D$-$V_{DS}$ curves along with an almost linear trend for all devices within the considered $V_{DS}$ range. The consistent behavior of the analyzed devices across voltages is also confirmed by Fig. 6(c) showing the drain-source resistance $R_{DS}$ as a function of $V_{DS}$. In order to quantify the variability in terms of conductivity, we extracted the $R_{DS}$ value from the linear fitting of the measured $I_D$-$V_{DS}$ characteristics (i.e., $R_{DSfit}$) for each device, as shown in Fig. 6(b). As a result of our extrapolation, Fig. 6(d) shows the $R_{DSfit}$ distribution with a mean value of 56.8 GΩ and a standard deviation of 16.6 GΩ, thus leading to $\sigma/\mu = 29.2\%$. With reference to the circuit shown in Fig. 1, such a mismatch in conductivity ensures random deviation of $V_{OUT}$ from the mid-supply point. Moreover, the similar bias dependence is highly beneficial to ensure adequate robustness against voltage variations. In other words, it ensures that the strength ratio between the two devices is well-maintained across voltages.

The variability of the device conductivity was also evaluated after subjecting the devices to different stabilization conditions. More specifically, for each device three different stabilization processes were applied in sequence and the output characteristics at zero-$V_{GS}$ were measured at the end of each stabilization process, i.e.: (i) after an initial stabilization phase with applying $V_{GS} = 0$ V for ~470 s, (ii) after a second stabilization phase with applying $V_{GS} = 35$ V for ~470 s, (iii) after two further stabilization phases, i.e., the first with $V_{GS} = 35$ V and the second with $V_{GS} = 0$ V, both for ~470 s. Fig. 7(a) shows the corresponding $I_D$-$V_{DS}$ curves related to the most conductive device among those analyzed (i.e., D2), compared to that measured without stabilization, whereas Fig. 7(b) reports the $R_{DSfit}$ as extracted from the $I_D$-$V_{DS}$ curves measured after different stabilization processes for all devices. From Fig. 7(a), again we can observe a linear trend of $I_D$ vs $V_{DS}$ regardless of the stabilization conditions. The measurements also reveal an opposite effect of the stabilization phase on the conductivity with respect to the applied voltage. Indeed, given the PMOS behavior of analyzed devices, stabilization with $V_{GS} = 35$ V leads to an increased conductivity (as already shown in Fig. 4) owing to charge detrapping phenomena at the oxide/channel interface. On the other hand, the stabilization at $V_{GS} = 0$ V leads to a decrease in the device conductivity likely due to charge trapping phenomena. However, such an effect is evident only in the third stabilization process, i.e., after first applying a stabilization at $V_{GS} = 35$ V which leads to a consequent initial charge detrapping. Therefore, the slight impact of the first stabilization process at $V_{GS} = 0$ V on the device conductivity may be ascribed to the high initial level of charge trapping in all analyzed devices. From Fig. 7(b), we can also observe that all devices reach an intermediate value of $R_{DSfit}$ when subjected to the third stabilization process, thus proving that the applied stabilization at $V_{GS} = 0$ V did not allow a complete recovery of the initial conductivity. Data shown in Fig. 7(b) again denote similar behavior across devices. It is also worth noting that, although the stabilization leads to a change in the device conductivity, a similar level of variability is observed in all cases (with $\sigma/\mu$ in the order of 30%). Moreover, the strength ratio between different devices is only slightly affected by the stabilization process. According to this observation, we expect the $V_{OUT}$ of the PUF circuit in Fig. 1 to be not affected by the trapping state at the oxide/channel interface.

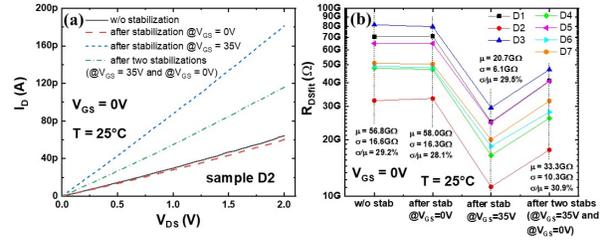

Fig. 7. Output characteristics at room temperature and zero-$V_{GS}$ under different stabilization conditions: (a) $I_D$-$V_{DS}$ curves of the most conductive device, i.e., D2 and (b) $R_{DSfit}$ for all tested devices.

## 4. Measurements across a temperature range

The electrical characteristics of DUTs were also evaluated across a temperature range. More specifically, the analysis was performed for increasing temperatures (i.e., from 25 °C up to 100 °C with a step of 25 °C).

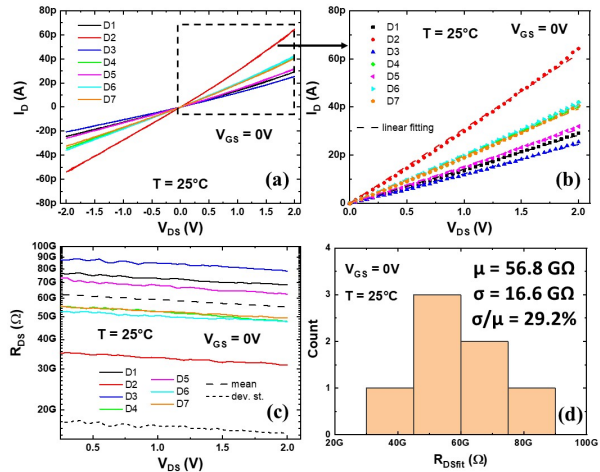

Fig. 6. Output characteristics of seven devices at 25 °C and zero-$V_{GS}$ without any initial stabilization: (a) $I_D$-$V_{DS}$ curves, (b) detail of $I_D$-$V_{DS}$ curves for $V_{DS}$ between 0 V and 2 V along with linear fitting, (c) $R_{DS}$ versus $V_{DS}$, (d) histogram of the fitting $R_{DS}$, i.e., $R_{DSfit}$, as extracted from the slope of linear fitting shown in (b).

Moreover, once the temperature was modified, a waiting time of a few hours was considered before performing *I-V* measurements given the relatively slow stabilization process in the analyzed devices.

Fig. 8(a) shows the $I_D$-$V_{GS}$ curves at $V_{DS}$ = 2 V and $T$ = 100 °C for all devices, measured after an initial stabilization with applying $V_{GS}$ = 35 V for ~470 s, whereas Fig. 8(b) reports the corresponding histogram of the hysteresis-induced $\Delta V_{GS}$ shift. When comparing these results with those measured at 25 °C (see Fig. 5), we can observe both increased conductivity and hysteresis at high temperatures, while again exhibiting an extremely low device-to-device variability in terms of hysteresis.

Measurements across temperatures were also performed for the output characteristics of DUTs. Figs. 9(a)-(c) summarize such results referred to the $I_D$-$V_{DS}$ curves measured at zero-$V_{GS}$ after an initial stabilization with applying $V_{GS}$ = 0 V for about 470 s. Fig. 9(a) compares the output characteristics of the most conductive device (i.e., D2) at different temperatures. Here, we can appreciate a strong dependence of the conductivity on the temperature. In particular, the performed characterization revealed an exponential-like increase of the current with increasing temperature. Such an effect could be attributed to an increase in the carrier concentration and/or a considerable impact of the Schottky barrier junctions formed at the source and drain contacts [11, 12]. Also, we can presume that the major sources of the observed variability in the electrical properties of DUTs are primarily ascribed to possible variations in structural and chemical features of the $WS_2$ film across the sample, which in turn could impact the channel resistivity as well as the potential barrier height at the metal contact/channel interfaces. The increasing trend of the conductivity for higher temperatures was consistently observed across all the devices, as given by Fig. 9(b) reporting the $R_{DSfit}$ (as extracted from the linear fitting of the $I_D$-$V_{DS}$ curves measured at different temperatures) as a function of temperature for all analyzed devices. The same figure shows the linear fitting of the $R_{DSfit}$-$T$ curves in a semilog scale, which allows effectively comparing the temperature dependence of the different devices. In this regard, Fig. 9(c) illustrates the histogram of the temperature coefficient $k_T$ corresponding to the (negative) slope of the fitting lines shown in Fig. 9(b). From this figure, the analyzed devices exhibit a quite similar temperature dependence, as proven by a variability in terms of $k_T$ equal to only 2.9%. Indeed, from Fig. 9(b), the relative strength ratio between different devices is quite well-maintained across temperatures, unless the cases with very small conductivity mismatch (e.g., D1 vs. D5 and D4 vs. D6). Such outcome is again beneficial for the reference circuit shown in Fig. 1, since a similar temperature dependence across devices suggests a limited chance of flipping the PUF response under temperature variations.

## 5. Conclusion

In this work, we experimentally investigated the variability of $WS_2$-based back-gated devices to assess their suitability for hardware security primitives such as PUFs. The electrical characteristics were evaluated across voltages and temperatures, as well as after applying different stabilization processes to the devices. Measurements demonstrated noticeable device-to-device variability in terms of conductivity (with $\sigma/\mu$ in the order of 30%). The devices also exhibited comparable bias and temperature dependences, as well as similar response under different stabilization conditions. Therefore, though our analysis was performed on a limited set of seven devices, overall results preliminarily prove the potential of such a technology to be exploited for implementing PUF solutions, while ensuring adequate randomness and robustness against environmental variations.

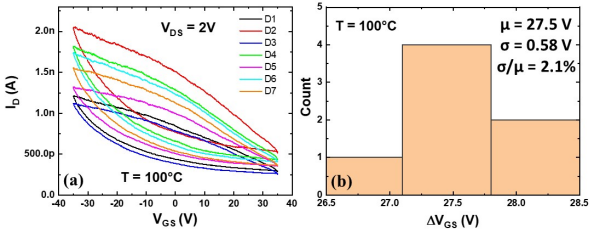

**Fig. 8.** (a) $I_D$-$V_{GS}$ curves at $V_{DS}$ = 2 V after stabilization with applying $V_{GS}$ = 35 V and (b) histogram of the hysteresis-induced $\Delta V_{GS}$ shift across seven devices at $T$ = 100 °C.

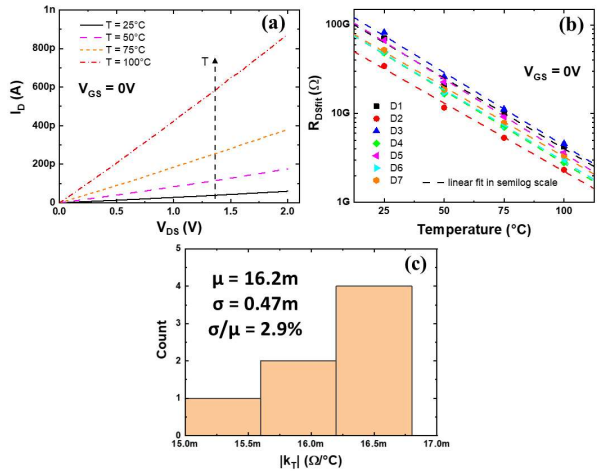

**Fig. 9.** (a) $I_D$-$V_{DS}$ curves of the most conductive device (i.e., D2) at zero-$V_{GS}$ across temperatures, as obtained after an initial stabilization with applying $V_{GS}$ = 0 V for about 470 s, (b) $R_{DSfit}$ versus temperature along with its linear fitting in a semilog scale across seven device, (c) histogram of the absolute value of the temperature coefficient $k_T$ as extracted from the slope of the linear fitting of $R_{DSfit}$-$T$ curves shown in (b).

## Acknowledgment

This work was partially supported by the European Union's Horizon 2020 project ASCENT+ (grant agreement no 871130), by Science Foundation Ireland (SFI-12/RC/2278_P2) and by the Italian MUR through the PRIN project FIVE2D (Contract No. 2017SRYEJH_001).